\documentclass[runningheads]{llncs}
\usepackage[T1]{fontenc}
\usepackage{amsmath,amsfonts}
\usepackage{algorithmic}
\usepackage{algorithm}
\usepackage{array}
\usepackage[caption=false,font=normalsize,labelfont=sf,textfont=sf]{subfig}
\usepackage{textcomp}
\usepackage{stfloats}
\usepackage{url}
\usepackage{verbatim}
\usepackage{multirow}
\usepackage{multicol}
\usepackage{mathptmx}
\usepackage[table]{xcolor} 
\usepackage{graphicx}
\usepackage{cite}
\usepackage{booktabs}

\begin{document}
\title{LSEG: A Lightweight and Secure Key Exchange Protocol for Smart Grid Communication}
%
%
\author{Amna Zafar \and
Muhammad Asfand Hafeez \and
Arslan Munir}

\institute{Florida Atlantic University, Florida, USA }
\maketitle              
\begin{abstract}
The increasing deployment of the Internet of Things (IoT) edge devices in modern smart grid environments requires secure and efficient communication protocols specifically designed for resource-constrained environments. However, most existing authentication schemes either impose excessive computational overhead or lack robustness against advanced cyber threats, making them unsuitable for resource-limited smart grid deployments. To address these limitations, this paper proposes a lightweight authentication and secure key exchange protocol for smart Grid (LSEG) environments. The proposed LSEG protocol utilizes a unified elliptic curve key pair, enabled by birational mapping between Ed25519 and Curve25519, for sign and key exchange. Initial keys are derived using the hash-based message authentication code (HMAC)-based key derivation function (HKDF), while ephemeral key pairs, generated through the Elliptic Curve Diffie–Hellman Ephemeral (ECDHE), are used in each session to ensure forward secrecy. Session communication is protected using ASCON128a, a lightweight, NIST-standardized, authenticated encryption algorithm. Formal security proofs in the random oracle model validate the security properties of the LSEG, including mutual authentication, forward secrecy, and resistance to impersonation, replay, and man-in-the-middle attacks. Experimental results on both Raspberry Pi and Intel Core i9-based systems demonstrate practical efficiency, achieving execution times under 5.5 milliseconds on embedded hardware and a communication cost of only 1024 bits for the protocol's message exchanges. The results demonstrate that the LSEG effectively balances security, efficiency, and compliance, making it a scalable solution for secure communication in smart grid infrastructures.

\end{abstract}
\section{Introduction}
The proliferation of Internet of Things (IoT) edge devices \cite{MunirEdgeAIAESM2021} within modern smart grid infrastructures is transforming the monitoring, management, and distribution of energy. These interconnected environments consist of thousands of sensors, smart meters, and control units that continuously exchange critical data over heterogeneous and often unsecured communication networks. Ensuring the security of this data exchange is essential, as any compromise can lead to severe disruptions in service availability, data confidentiality, or operational integrity \cite{mcdaniel2009security, karale2021challenges}. Consequently, secure communication protocols specialized to the unique characteristics of smart grid systems have become an area of intense research interest \cite{zyskind2015decentralizing}. 

Despite the availability of traditional cryptographic solutions offering strong security assurances, their applicability in smart grid environments is limited. The key challenge in smart grid environments is to strike a balance between the need for robust security and the limited computational resources available at edge devices. Many such traditional protocols are built on RSA or classical elliptic curve cryptography (ECC)~\cite{rivest1978method, koblitz1987elliptic}, which are computationally expensive. These protocols often require large public keys, intensive computations, and complex algorithms, resulting in significant latency and power consumption. This imposes significant computational overhead, excessive energy consumption, and substantial memory requirements, making them unsuitable for deployment in embedded field devices with constrained resources, such as low power microcontrollers and battery-operated nodes \cite{sandeep2006elliptic}. In addition, full public-key infrastructure (PKI) implementations, especially those relying on standard X.509 certificates~\cite{cooper2008internet}, exacerbate latency and communication overhead, often rendering secure handshakes impractical in time-sensitive environments.  

In addition to resource constraints, smart grid networks face challenges related to scalability and dynamic topologies. Devices may frequently join or leave the network, necessitating robust, decentralized, and scalable authentication mechanisms. Centralized key distribution or authentication architectures cannot meet the demand for low-latency, autonomous communication at scale, especially with low-cost devices deployed in geographically dispersed areas \cite{won2018decentralized}. This dynamic environment demands lightweight, decentralized cryptographic protocols that provide end-to-end security guarantees while maintaining computational and energy efficiency.

Several lightweight authentication protocols have been proposed in recent years to address security challenges in resource-constrained smart grid environments. For example, Tanveer et al. \cite{tanveer2023lacp} introduced an efficient lightweight scheme; however, it lacks mechanisms for mutual authentication and session key agreement, rendering it vulnerable to Denial-of-Service (DoS) attacks. Similarly, Aziz et al. \cite{aziz2018lightweight} proposed a scheme that achieves mutual authentication by combining static identifiers, timestamps, and random numbers using basic XOR and concatenation operations. This protocol enables session keys to be computed by an adversary from a single transcript, due to the static reuse of pseudonyms and long-term secret values, violating forward secrecy and session uniqueness. As a result, the protocol is susceptible to impersonation, man-in-the-middle (MitM), and replay attacks.

Many existing solutions reduce computational complexity at the expense of cryptographic robustness. Protocols that rely on symmetric key cryptography or pre-distributed secrets, such as ANSI C12.18, are inherently limited in scalability and security. Furthermore, several schemes do not support mutual authentication, lack forward secrecy, or are incompatible with standard certificate infrastructures. Only a few are supported by formal security proofs or validated through real-world embedded platform implementations. To overcome these limitations, this paper presents a Lightweight, certificate-based authentication and Secure key Exchange for smart Grid (LSEG). The proposed LSEG protocol is designed specifically for resource-constrained smart grid environments. LSEG is designed to support mutual authentication, forward secrecy, and secure session key establishment while minimizing memory usage, energy consumption, and communication overhead. The main contributions of this work are as follows:

\begin{enumerate}
    \item We propose a novel, lightweight key agreement protocol LSEG that unifies digital signature and key-exchange operations by using a bijective birational mapping between Ed25519 and X25519. This enables the reuse of a single 32-byte elliptic curve key pair for both operations, significantly reducing memory overhead and certificate size.
    \item The proposed protocol LSEG integrates ASCON128a, a lightweight NIST standard authenticated encryption scheme to secure communication, thereby ensuring data confidentiality, integrity, and resistance to nonce reuse, while maintaining low computational latency and energy consumption.
    \item A comprehensive security analysis under the random oracle model formally establishes the protocol's resistance to session key compromise, impersonation, and man-in-the-middle attacks. An informal analysis also addresses real-world threats, including replay, insider, and denial-of-service attacks.
    \item The protocol is implemented and experimentally evaluated on both Raspberry Pi and high-performance desktop systems. The results demonstrate its practical viability, with low execution time, minimal communication cost (1024 bits in two messages), and suitability for deployment in real-time, resource-constrained smart grid environments.
\end{enumerate}
The remainder of the paper is structured as follows. Section \ref{sec:back} discusses the background and related work. Section \ref{sec:protocl_design} details the proposed protocol. Section \ref{security} discusses security analysis. Section \ref{sec:results} presents experimental results and performance evaluation. Finally, Section \ref{sec:conclusion} concludes the paper and outlines future research directions.
\section{Background}\label{sec:back}

This section first reviews the cryptographic primitives that underlie the proposed protocol. Subsequently, it discusses the related research in the literature, highlighting important studies and their relevance to this work.

\subsection{Preliminaries}\label{sec:pre}
\subsubsection{Ed25519 Digital Signature:}
Ed25519 is a deterministic digital signature scheme introduced by Bernstein et al.~\cite{bernstein2012high}. It is a fixed parameter instantiation of the EdDSA signature scheme over the prime field \( \mathbb{F}_p \), where \( p = 2^{255} - 19 \). The underlying curve is the twisted Edwards form:

\begin{equation}
x^2 + y^2 = 1 + d x^2 y^2, \quad d = \frac{-121665}{121666} \pmod{p}
\end{equation}
A pre-defined base point \( B \) on the curve has a prime order;
\[
l = 2^{252} + 27742317777372353535851937790883648493
\]
where $l$ denotes the order of the subgroup generated by $B$. The group has a cofactor of $h=8$, indicating that the total number of points on the curve is given by $h\cdot l$. All computations are restricted to the subgroup of order \( l \), which provides approximately 128-bit classical security. The keys are deterministically derived from a 32-byte seed, producing 32-byte public keys and 64-byte signatures \( (R, S) \). Ed25519 leverages SHA-512 and ensures high efficiency and robustness based on the assumed hardness of the discrete logarithm problem over the specified group.

\subsubsection{X25519 Elliptic Curve Diffie–Hellman:}
X25519 is an elliptic curve Diffie-Hellman (ECDH) key exchange mechanism in Montgomery form defined over an identical field \( \mathbb{F}_p \), operating on the curve:

\begin{equation}  
M: y^2 = x^3 + 486662 x^2 + x
\end{equation}
The base point uses the $x$-coordinate \( u = 9 \). Scalar multiplication is performed using the 32-byte clamped private scalar $k$ with the Montgomery ladder, generating the public key \( X = [k]u \). Two parties with private scalars \( k_A \) and \( k_B \) compute the shared secret as:
\[
[k_A]X_B = [k_B]X_A
\]
where $X_A=[k_A]u$ and $X_B=[k_B]u$ denote the public keys of $A$ and $B$ parties, respectively. The public and private keys, as well as the resulting shared secret, are each 32 bytes in length. X25519 offers 128-bit classical security, thus avoiding cofactor-related vulnerabilities, making it suitable for constrained systems.

\subsubsection{HMAC-based Key Derivation Function (HKDF):}
HKDF is a key derivation function defined in RFC 5869 that transforms potentially weak input keying material (IKM) into cryptographically strong output keying material (OKM) using HMAC~\cite{krawczyk2010cryptographic}. Strong output keying material refers to key material that is computationally indistinguishable from random and suitable for direct use in cryptographic operations such as encryption or message authentication. HKDF operates in two phases: extract and expand.

\[
Extract: \mathsf{PRK} = \mathsf{HMAC}(\text{salt}, \text{IKM})
\]
where PRK denotes the pseudorandom key, an intermediate fixed-length key derived from the IKM and an optional salt.
\[
\begin{aligned}
\text{Expand:} \quad & T_0 = \emptyset, \\
& T_i = \mathsf{HMAC}(\mathsf{PRK},\, T_{i-1} \,\|\, \text{info} \,\|\, i), \\
& \text{for } 1 \leq i \leq N = \left\lceil \frac{L}{\ell_{\mathrm{hash}}} \right\rceil
\end{aligned}
\]
Here, info is optional context or application-specific information (e.g., a protocol identifier, session identifier, or role tag) used to bind the derived keys to a particular use case. L denotes the desired length of the output keying material in bytes, and $l_{hash}$ is the output length of the underlying hash function (e.g., 32 bytes for SHA-256).The final output is:
\[
\mathsf{OKM} = T_1 \| T_2 \| \dots \| T_N[0:L]
\]
Here, \( \ell_{hash} \) is the hash output length (e.g. 32 bytes for SHA-256). The security of HKDF relies on the pseudorandom function property of HMAC, ensuring domain separation,and in-distinguishability from random output, provided that sufficient entropy is present in the IKM.

\subsubsection{Birational Mapping Between Curves:}
Let \( X \) and \( Y \) be algebraic curves over a field \( k \). A birational map \( \varphi: X \to Y \) is a rational map for which there exists an inverse \( \psi: Y \to X \) such that:

\[
\psi \circ \varphi = \mathrm{id}_X, \quad \varphi \circ \psi = \mathrm{id}_Y
\]
where $\mathrm{id}_X$ and $\mathrm{id}_Y$ denote the identity maps in $X$ and $Y$, respectively. This implies that \( X \) and \( Y \) are isomorphic in dense open subsets and share isomorphic function fields, i.e., \( k(X) \cong k(Y) \)~\cite{hartshorne2013algebraic}, where $k(X)$ and $k(Y)$ denote the function fields of $X$ and $Y$. In this work, this property is used to map elliptic curve keys between the Edwards and Montgomery forms, enabling dual use of a single 32-byte key pair for both signature and key-exchange operations. This approach significantly reduces memory overhead and simplifies implementation.

\subsubsection{ASCON128a Authenticated Encryption:}

ASCON128a, selected by NIST in 2023, is a member of the ASCON family of lightweight authenticated encryption algorithms. It maintains a 320-bit internal state, divided into a 128-bit rate and a 192-bit capacity. Initialization incorporates a 128-bit key and nonce using a 12-round permutation \( P_{12} \).
In the authenticated encryption with associated data (AEAD) setting, associated data refers to additional information that is authenticated but not encrypted. In this work, the associated data consist of the protocol session identifiers (total length: 128 bits), ensuring that messages are bound to the correct protocol instance and preventing replay or cross-protocol attacks. The associated data is absorbed into the state, and plaintext is processed in blocks using an 8-round permutation \( P_{8} \),producing a 128-bit authentication tag. ASCON128a offers 128-bit confidentiality and integrity, assuming nonces are not reused. Its lightweight operations, primarily XOR, AND, and S-box substitutions, make it suitable for constrained embedded devices while offering high throughput on modern CPUs.

\subsection{Literature Review}\label{sec:leterature}
Modern smart grids comprise millions of embedded nodes (e.g., smart meters, phasor measurement units) that operate networks outside direct utility control. Achieving server-grade cryptographic security guarantees on devices with only kilobytes of memory and milliwatts of power (e.g., battery powered smart meters) has become a defining challenge for power system engineers.

Smart grid security research has advanced rapidly over the past fifteen years, yet field devices remain the most common point of failure. Early deployments added cryptographic wrappers to existing supervisory control protocols. DNP3 Secure Authentication appends message authentication codes to legacy DNP3 traffic \cite{gilchrist2008secure}, and IEC 62351‑3 specifies a Transport Layer Security (TLS) profile for IEC 60870‑5‑104 frames \cite{iec200762351}. Large‑scale residential roll‑outs rely on DLMS/COSEM with mutual X.509 authentication defined by the DLMS User Association \cite{dlms2019green}, while IEEE 2030.5 (Smart Energy Profile 2.0) requires TLS for home energy gateways \cite{ieee2018ieee}. Empirical studies show that a full TLS handshake can increase connection latency by up to three times and exhaust the limited RAM of 32‑bit microcontrollers. As a result,vendors often disable certificate validation or revert to preshared keys, leaving well-documented security gaps.

The December 2015 Ukraine blackout showed how an enterprise breach of corporate IT networks of regional power distribution companies can be leveraged to pivot to remote terminal units and send unauthenticated breaker commands. This attack interrupted service to approximately 225,000 customers \cite{case2016analysis}. Attack graph analysis supports this observation. Ten et al. found that compromising a single unprotected field device markedly lowers the effort required to disrupt an entire feeder \cite{ten2010cybersecurity}. Similarly, Sridhar et al. mapped cyberphysical dependencies and concluded that intelligent electronic devices remain the most exposed assets in the grid \cite{sridhar2011cyber}.

To maintain strong authentication without overwhelming low-power hardware, recent research has developed authentication schemes specifically optimized for smart grid environments. Mahmood et al. introduced a lightweight message authentication scheme that combines hash functions and XOR operations, reducing computational overhead for resource‑limited devices \cite{mahmood2016lightweight}. Li et al. proposed a lightweight protocol based on ECC that reduces both storage and computational costs while providing mutual authentication and forward secrecy between smart meters and gateway nodes \cite{li2013enhanced}. Gope and Sikdar presented a Privacy Preservation Framework based on physically unclonable functions (PUF) that offers hardware-based security with minimal processing, making it suitable for constrained smart grid infrastructures \cite{gope2018lightweight}. In general, these studies demonstrate that practical and secure authentication can be tailored to meet the stringent performance requirements of smart grid devices.
Beyond hash-based and PUF-oriented methods, researchers are now incorporating modern ECC primitives to enhance efficiency and security. Challa et al. designed an ECC key agreement protocol for smart grid environments that reduces computational and communication overhead while resisting impersonation and replay attacks \cite{challa2018efficient}. Abbasinezhad et al. proposed an ECC‑based anonymous authentication scheme for vehicle‑to‑grid links, reducing computation while guaranteeing user anonymity and non-traceability \cite{abbasinezhad2017ultra}. Baghestani et al. later combined ECC with hash‑based authentication in a smart meter protocol formally verified with Automated Validation of Internet Security Protocols and Applications (AVISPA), demonstrating resistance to common attacks and achieving low latency and energy consumption on constrained devices \cite{baghestani2022lightweight}. These results confirm that ECC approaches can deliver significant security and performance gains for smart grid authentication.

Existing protocols typically achieve low overhead or strong security, but few deliver certificate-based trust, key reuse, and lightweight authenticated encryption within the 8-kilobyte memory budget of field devices. To address this limitation, we propose a protocol that utilizes a single elliptic-curve key pair for digital signatures and key exchange, maintains X.509 compatibility through compact certificate processing, and secures data with ASCON128a, using keys derived from an HMAC-based key-derivation function. Our formal analysis confirms that the proposed LSEG resists replay and impersonation attacks while satisfying the latency, memory, and energy constraints of real-time smart grid deployments.

\section{Protocol Design} \label{sec:protocl_design}
\begin{figure*}[t]
    \centering
    \includegraphics[width=0.8\textwidth]{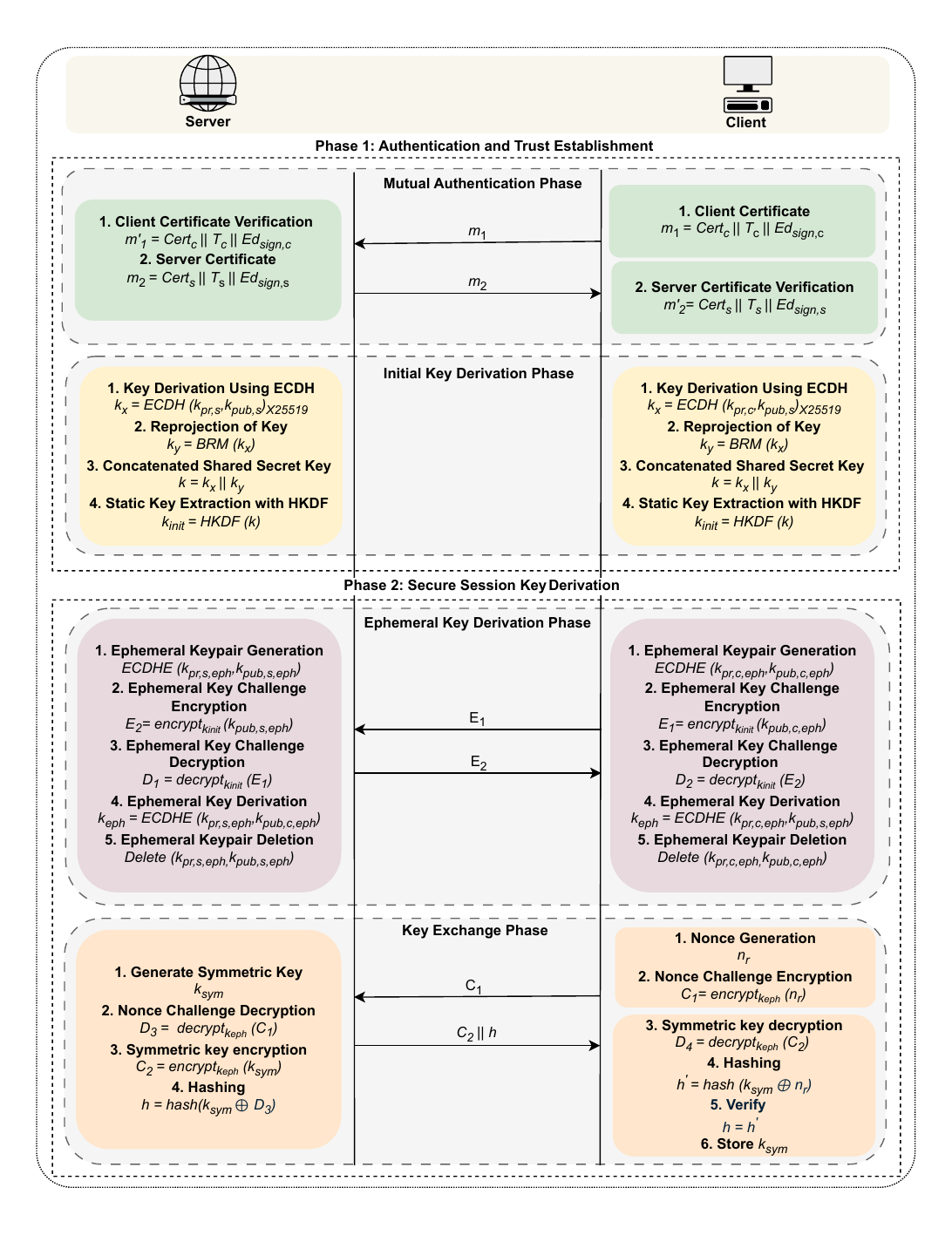}
    \caption{Framework of the proposed lightweight protocol LSEG.}
    \label{fig_1}
\end{figure*}

This section describes the proposed lightweight authentication and the secure communication protocol. LSEG consists of two distinct phases: authentication and trust establishment, and secure session key derivation. The notation used throughout is defined separately in Table~\ref{tab:notations}, and Figure~\ref{fig_1} illustrates the framework of the proposed protocol.

\subsection{Authentication and Trust Establishment}The phase 1 of the protocol is composed of two subsections, Mutual Authentication and Key Derivation.In this phase, both parties establish trust by validating credentials of each other and subsequently derive an initial shared key. This phase is performed once to establish trust and does not need to be repeated in subsequent sessions between the same peers.

\subsubsection{Mutual authentication phase:}
The mutual authentication phase ensures that both the client and the server verify the identity of each other before establishing further secure communications. For identity verification, standard X.509 certificates are used, which securely bind each participant's public key to their identity. The client and server certificates are represented as $Cert_c$ and $Cert_s$, respectively. The client initiates authentication by sending a message containing its certificate, a fresh timestamp $T_c$, and an Ed25519 digital signature created using its private key $k_{pr,c}$:
\begin{equation}
  m_1 = \{Cert_c, T_c, Ed_{sig,c}(Cert_c\| T_c)\}
\end{equation}

Upon receiving $m_1$, the server performs three verifications: the validity of the client certificate, the freshness of the timestamp to prevent replay attacks, and the correctness of the signature using the public key of the client. If verification passes, the server sends its certificate, fresh timestamp $T_s$, and signature computed similarly to the client:

\begin{equation}
  m_2 = \{Cert_s, T_s, Ed_{sig,s}(Cert_s\| T_s)\}
\end{equation}

The client performs the three verification steps. After successful verification of both messages, mutual authentication is completed.

\begin{table}[ht]
\centering
\caption{Notations used in the proposed protocol}
\label{tab:notations}
\begin{tabular}{p{0.20\linewidth} p{0.65\linewidth}}
\textbf{Symbol} & \textbf{Description} \\
\toprule

$Cert_c$ & Certificate for client \\
$Cert_s$ & Certificate for server \\
$T_c$ & Timestamp for client \\
$T_s$ & Timestamp for server \\
$k_{pr,c}, k_{pub,c}$ & Private and public key of client \\
$k_{pr,s}, k_{pub,s}$ & Private and public key of server \\
$k_x$ & Key generated using X25519 \\
$k_y$ & Key generated using reprojection \\
$k$ & Concatenation $k_x || k_y$ \\
$k_{init}$ & Initial key \\
$k_{pub,c,eph},k_{pr,c,eph}$ & Ephemeral key pair for client \\
$k_{pub,s,eph},k_{pr,s,eph}$ & Ephemeral key pair for server \\
$k_{eph}$ & Ephemeral key \\
$k_{sym}$ & Symmetric session key \\ 
$n_r$ & Nonce \\
$h, h'$ & Hashes of nonce \\
$C_1$ & Encrypted nonce \\
$C_2$ & Encrypted nonce with hash \\
$D_1, D_2$ & Decryption of ephemeral key challenge \\
$D_3$ & Decryption of encrypted nonce \\
$D_4$ & Decryption of encrypted symmetric key \\
$Ed_{sig}$ & Ed25519 signature algorithm \\
$BRM$ & Birational mapping \\
$HKDF$ & HMAC-based key derivation function \\ 
\bottomrule
\end{tabular}
\end{table}

\subsubsection{Initial key derivation phase:}

After authentication, both sides compute a shared symmetric initial session key. The initial key derivation phase employs an ECDH using the X25519 algorithm, providing forward secrecy with a minimal computational load suitable for constrained devices. The client computes a secret key $k_x$ using its private key and the public key of the server. Similarly, the server computes a secret $k_x$ with its private key and public key of the client:

\begin{equation}
\begin{aligned}
  k_x &= {ECDH}(k_{pr,c}, k_{pub,s})_{X25519} \\
  k_x &= {ECDH}(k_{pr,s}, k_{pub,c})_{X25519}
\end{aligned}
\end{equation}
X25519 produces results on a Montgomery-form curve. To align this with Edwards form, a birational mapping converts the Montgomery curve x-coordinate ($x$) into an Edwards curve y-coordinate ($y$):

\begin{equation}
  y = \frac{x - 1}{x + 1}
\end{equation}
The conversion of the Montgomery curve into the Edwards curve is denoted by $k_y$. This mapping ensures compatibility and cryptographic correctness. Both parties apply this mapping to their computed secrets. The mapped results $k_x$ and $k_y$ are concatenated $k=k_x||k_y$. The initial key $k_{init}$ is derived using HKDF with HMAC-SHA256 and a non-zero salt defined as SHA-256 $(k_{pub,c}, k_{pub,s})$;

\begin{equation}
  k_{init} = {HKDF}(k)
\end{equation}
The resulting key $k_{init}$ is uniformly random and is used for secure symmetric encryption in the next key exchange phase.
\subsection{Secure Session Key Derivation}
The second phase of the proposed protocol includes
two sub phases; ephemeral key derivation and key exchange phase. Since trust is established in Phase 1, this phase is executed in subsequent sessions to derive fresh session keys to ensure forward secrecy. This session key is then used to secure all subsequent communications.

\subsubsection{Ephemeral key derivation phase:}
In the ephemeral key derivation phase, a session key is derived to further enhance the confidentiality of the communication channel and provide forward secrecy. This phase involves generating fresh, short-lived ECDHE key pairs on both the client and the server sides. These ephemeral keys are used to derive a session key that is unlinkable to long-term credentials and is valid only for the current session. Both the client and the server side generate their ephemeral keys. To ensure the confidentiality of the exchanged ephemeral public keys, they are encrypted using the initial key $k_init$ derived in the key derivation phase. Each ASCON128a encryption includes associated data (AD) that binds ciphertexts to the communicating parties. Assoicated data (AD) is defined as the concatenation of the client and server identifiers:

\begin{equation}
\begin{aligned}
    AD = ID_c || ID_s
\end{aligned}
\end{equation}
The encrypted messages, known as ephemeral key challenges, are computed and exchanged as follows:
\begin{equation}
\begin{aligned}
    E_1= \operatorname{AsconEnc}_{k_{init}}(AD,k_{pub,c,eph})\\
    E_2= \operatorname{AsconEnc}_{k_{init}}(AD,k_{pub,s,eph})
\end{aligned}
\end{equation}
Upon receiving $E_1$ and $E_2$, both sides decrypt the messages using the initial key$k_{init}$ to obtain ephemeral public keys of each other. Using the received ephemeral public key and their own ephemeral private key, both entities compute a new ephemeral shared session key using ECDHE.
\begin{equation}
\begin{aligned}
    k_{eph}=ECDHE(k_{pr,s,eph},k_{pub,c,eph}) \\
    k_{eph}=ECDHE(k_{pr,c,eph},k_{pub,s,eph})
\end{aligned}
\end{equation}
This ensures that both the client and the server derive an identical ephemeral key $k_{eph}$
without ever exposing their private key material. Immediately after the key agreement, both sides securely erase their ephemeral key pairs from memory. This prevents compromise of long-term session keys, even if ephemeral material is later exposed. The ephemeral session key is then used in the subsequent key exchange phase for encrypted communication.

\subsubsection{Key exchange phase:}

In the key exchange phase, the client generates a fresh \(128\)-bit nonce \(n_r\) and protects it with authenticated encryption using the ephemeral key \(k_{eph}\):
\begin{equation}
  C_1 = \operatorname{AsconEnc}_{k_{eph}}(n_r).   
\end{equation}
Since ASCON128a is an AEAD scheme, this step simultaneously ensures the confidentiality and integrity of \(n_r\). Although ASCON128a provides confidentiality and integrity, the additional hash of the nonce serves as an explicit key-confirmation step linked to the ephemeral Diffie–Hellman output, strengthening guarantees of forward secrecy. It also enables quick rejection of forged or replayed messages prior to full decryption, reducing the risk of DoS on constrained smart grid devices. The client sends the encrypted nonce \((C_1 )\) to the server. Upon receipt, the server decrypts the ciphertext,
\begin{equation}
      D_3 = \operatorname{AsconDec}_{k_{eph}}(C_1),
\end{equation}
 The server generates a fresh symmetric key \(k_{sym}\) of 128 bits, encrypts it under \(k_{eph}\),
\begin{equation}
  C_2 = \operatorname{AsconEnc}_{k_{eph}}(k_{sym} ),  
\end{equation}
The server, after recovering $n_r$, computes a verification hash:
\begin{equation}
  h = \operatorname{Hash}(k_{sym} \oplus D_3).
\end{equation}
and sends the pair $(C_2, h)$ to the client. Upon receiving $(C_2, h)$, the client decrypts $C_2$ to obtain $k_{sym}$ and recomputes the hash for verification;
\begin{equation}
  h' = \operatorname{Hash}(k_{sym} \oplus n_r).
\end{equation}
If $h' = h$, the session key $k_{sym}$ is accepted. This ensures that the key is fresh, bound to the original nonce, and mutually confirmed by both parties. This exchange verifies that both parties possess $k_{eph}$ and securely derive the session key $k_{sym}$, which is then stored for subsequent communication.

All subsequent application data are encrypted and authenticated with ASCON128a under \(k_{sym}\), thus preserving confidentiality and integrity at low computational cost.

\section{Security Analysis}\label{security} 
In this section, we discuss the formal and informal security analysis of the proposed scheme. Later, we will also discuss the comparative analysis of the proposed scheme with the existing state-of-the-art work.

\subsection{Formal Analysis}
The goal of this proof is to show that the proposed lightweight public-key authentication and key-agreement protocol achieves session key indistinguishability in the Real-or-Random (ROR) model. The analysis is performed in the Random Oracle Model (ROM), where all hash functions, including those used inside HKDF and for key confirmation, are modeled as random oracles. Specifically, the objective is to prove that a probabilistic polynomial-time adversary $\mathcal{A}$ cannot distinguish between a real session key and a random key, except with negligible advantage, under the stated cryptographic assumptions.  

We denote the security parameter by $\lambda$. Let $\Pi$ be the protocol and let $\Pr[\mathsf{G}_i=1]$ represent the probability that $\mathcal{A}$ correctly guesses the challenge bit in the game $\mathsf{G}_i$. The proof proceeds through a series of games, each bounding the advantage of the adversary under a well-defined assumption.  
\begin{table*}[!t]
\centering
\caption{Sequence of games in the AKE security proof of the proposed protocol.}
\begin{tabular}{p{0.3\linewidth} p{0.65\linewidth}}
\textbf{Games} & \textbf{Description} \\
\toprule

\textbf{Game 0} (Real Execution) &
Adversary interacts with the actual protocol: Ed25519/X.509 mutual authentication with timestamps, static DH using X25519 to derive $k_x$, birational mapping to $k_y$, and HKDF to compute $k_{init}$. This initial key protects the transport of fresh ephemeral public keys using ASCON128a. Both parties generate ephemeral X25519 keys, exchange them under $k_{init}$, and derive $k_{eph}$. The client proves knowledge of $k_{eph}$ by sending an encrypted nonce, and the server responds with $k_{sym}$ encrypted under $k_{eph}$, along with a hash $h = Hash(k_{sym} \oplus n_r)$. The adversary attempts to distinguish $k_{sym}$ from random. \\[2ex]

\textbf{Game 1} (Random Oracle Simulation) &
All hash functions, including those in HKDF and key confirmation, are replaced by random oracles. This modification is in-distinguishable to the adversary except with probability at most:
$\bigl|\Pr[\mathsf{G}_0 = 1] - \Pr[\mathsf{G}_1 = 1]\bigr| \leq \tfrac{q_H^2}{2^\lambda}$. \\[2ex]

\textbf{Game 2} (Signature Forgery) &
The adversary may attempt to forge an Ed25519 signature to impersonate a party. By the EUF-CMA security of Ed25519, this happens with probability at most:
$\bigl|\Pr[\mathsf{G}_1 = 1] - \Pr[\mathsf{G}_2 = 1]\bigr| \leq \epsilon_{Sig}$. \\[2ex]

\textbf{Game 3} (Replace Static Diffie Hellman Secret) &
The static shared secret $(k_x, k_y, k_{init})$ is replaced with a uniform random string. By the CDH assumption on X25519, the adversary cannot distinguish this replacement:
$\bigl|\Pr[\mathsf{G}_2 = 1] - \Pr[\mathsf{G}_3 = 1]\bigr| \leq \epsilon_{CDH}$. Since $k_{init}$ only protects the ephemeral key exchange, this does not affect $k_{sym}$. \\[2ex]

\textbf{Game 4} (Replace Ephemeral Diffie Hellman Secret) &
The ephemeral shared secret $k_{eph}$ is replaced by a uniform random string. By the CDH assumption on Curve25519, this change is indistinguishable except with probability:
$\bigl|\Pr[\mathsf{G}_3 = 1] - \Pr[\mathsf{G}_4 = 1]\bigr| \leq \epsilon_{CDH-eph}$. \\[2ex]

\textbf{Game 5} (Replace ASCON Encryption) &
All ASCON128a ciphertexts (ephemeral public keys, encrypted nonce, and encrypted session key) are replaced with random strings. By the IND-CPA security of ASCON:
$\bigl|\Pr[\mathsf{G}_4 = 1] - \Pr[\mathsf{G}_5 = 1]\bigr| \leq \epsilon_{Ascon}$. \\[2ex]

\textbf{Game 6} (Hash Based Key Confirmation) &
Session acceptance requires both sides to compute and verify $h = \mathsf{Hash}(k_{sym} \oplus n_r)$. Any mismatch aborts the session. The advantage of adversary is bounded by the collision resistance of the hash:
$\bigl|\Pr[\mathsf{G}_5 = 1] - \Pr[\mathsf{G}_6 = 1]\bigr| \leq \epsilon_{Hash}$. At this point, $k_{sym}$ is independent of the view of adversary, so
$\Pr[\mathsf{G}_6 = 1] = \tfrac12$. \\

\bottomrule
\end{tabular}
\label{fig:game_proof}
\end{table*}

\textbf{Game 0} (Real Execution):
The adversary observes all protocol messages, including Ed25519-based authentication, X25519 DH key exchange, ASCON128a ciphertexts, and the final hash-based key confirmation.  

\textbf{Game 1} (Random Oracle):  
All hash functions are modeled as random oracles. This is indistinguishable from $\mathcal{A}$, except for probability at most $q_H^2/2^\lambda$, where $q_H$ is the number of hash queries. 

\textbf{Game 2} (Signature Forgery):
The adversary may attempt to forge a valid Ed25519 signature. By the EUF-CMA security of Ed25519, the probability of success is bounded by $\epsilon_{Sig}$.  

\textbf{Game 3 } (Static Diffie–Hellman Randomization): 
The static shared secret~$(k_x, k_y)$ and the derived key~$k_{init}$ are replaced by uniform random strings.  By the Computational Diffie–Hellman (CDH) assumption on Curve25519, this substitution is indistinguishable to~$\mathcal{A}$ except with probability~$\epsilon_{CDH}$:
\[
\bigl|\Pr[\mathsf{G}_2=1]-\Pr[\mathsf{G}_3=1]\bigr|\le \epsilon_{CDH}.
\]

\textbf{Game 4} (Ephemeral Diffie–Hellman Randomization):
The ephemeral shared secret~$k_{eph}$ is now replaced with a random string.  Under the CDH assumption for the ephemeral X25519 keys, the adversary cannot detect this modification with advantage greater than~$\epsilon_{CDH-eph}$:
\[
\bigl|\Pr[\mathsf{G}_3=1]-\Pr[\mathsf{G}_4=1]\bigr|\le \epsilon_{CDH-eph}.
\]

\textbf{Game 5} (Hash Based Key Confirmation):  
Both parties accept the session key only if they compute the same hash value $h = Hash(k_{sym} \oplus n_r)$. Any mismatch causes the session to abort. The adversary cannot succeed unless it breaks the hash collision resistance or learns both $k_{sym}$ and the fresh nonce $n_r$, with probability bounded by $\epsilon_{Hash}$. At the end of Game 5, the session key is independent of the adversary’s view, and $\Pr[\mathsf{G}_5=1] = 1/2$. The overall advantage of the adversary in the real protocol is therefore bounded by:  
\begin{equation}
  \mathsf{Adv}_{\mathcal{A}}^{\mathsf{AKE}} 
  \leq \frac{q_H^2}{2^\lambda} + \epsilon_{Sig} + \epsilon_{\mathsf{CDH}} + \epsilon_{Ascon} + \epsilon_{Hash}.
\end{equation}
Thus, under the assumptions of Ed25519 signature security (EUF-CMA), the hardness of CDH on Curve25519, the IND-CPA security of ASCON128a, and the collision resistance of the hash function, the proposed protocol achieves session key in-distinguishability in the ROR model.

\subsection{Informal Analysis}
\subsubsection{Mutual Authentication:}The protocol achieves mutual authentication by pairing an Ed25519 signature with a fresh timestamp. The client first sends \(m_1 = ({Cert}_c, T_c, \sigma_c)\) where \(\sigma_c =EdSig_{k_{{pr},c}}({Cert}_c \parallel T_c)\). The server validates \({Cert}_c\), checks that the timestamp \(T_c\) is within the allowed clock skew \(\Delta t\), and verifies the signature with \(k_{{pub},c}\). If all checks are successful, the server responds with \(m_2 = ({Cert}_s, T_s, \sigma_s)\) where \(\sigma_s = EdSig_{k_{{pr},s}}({Cert}_s \parallel T_s)\). The client then performs the same certificate, timestamp and signature verifications using \(k_{{pub},s}\). Because only legitimate holders of \(k_{{pr},c}\) and \(k_{{pr},s}\) can produce valid signatures, and stale timestamps are rejected, an attacker cannot forge, modify, or replay either message. Therefore, both parties confirm the identity of each other and the freshness of the session.
\subsubsection{Session Key Agreement:}The client and server generate an X25519 static key pair at the start of the session. This key is not used for application data, but only to encrypt and protect the exchange of fresh ephemeral public keys. From these, both sides compute an ephemeral shared secret key. The client proves possession of the ephemeral key by sending an encrypted nonce, and the server replies with the final symmetric key, also encrypted under $k_{eph}$. It ensures that the true session key depends only on a short-lived ephemeral key, while preventing tampering during their exchange. To provide explicit key confirmation, both sides compute and compare a hash that binds the symmetric session key $k_{sym}$ to the nonce $n_r$ generated by the client. This step guarantees that the session key is fresh, prevents replay or desynchronization attacks, and ensures that both parties possess the same key before it is accepted for subsequent communication. 

\subsubsection{Replay Attack Resistance:}The proposed protocol prevents replay attacks by binding each handshake to fresh values. Authenticated messages include new timestamps signed with Ed25519, so stale packets are rejected. In the key exchange phase, the client encrypts a random nonce under the ephemeral key $k_{eph}$; the server decrypts it and responds with the final session key $k_{sym}$, also encrypted under$k_{eph}$.To provide explicit key confirmation, both sides compute and compare a hash that binds $k_{sym}$ to the nonce $n_r$ generated by the client. Since both the ephemeral key and the nonce are unique to each run, replayed messages fail verification and are discarded before a session is established.

\subsubsection{Man-in-the-Middle Attack Resistance:}The proposed protocol protects against man-in-the-middle attacks by combining certificate-based signatures, fresh timestamps, and authenticated Diffie–Hellman exchanges. Each peer signs its certificate and timestamp with Ed25519, so any modification breaks the signature. After authentication, both sides derive an initial key $k_{init}$ from X25519, which is used only to protect the exchange of ephemeral public keys. The ephemeral key $k_{eph}$ then depends on both private ephemeral values, making it infeasible for an interceptor to compute. The client proves knowledge of $k_{eph}$ by sending an encrypted nonce, and the server returns the final session key $k_{sym}$ with the same key. Any attempt by an attacker to insert its own key material fails decryption or tag verification, exposing the attack.

\subsubsection{Forward Secrecy:}
Forward secrecy is achieved through the use of ephemeral key pairs. After phase 1, both the client and the server generate Curve25519 keys and exchange them under the protection of $k_{init}$. The resulting ephemeral secret $k_{eph}$ forms the basis of the final session key $k_{sym}$. To guarantee explicit key confirmation and freshness, both sides compute and compare a hash that binds $k_{sym}$ to the nonce $n_r$ generated by the client. Since ephemeral keys are securely erased after exchange, a later compromise of the long-term signing keys or even $k_{sym}$ cannot reveal the keys of previous sessions. Each session is therefore cryptographically independent, ensuring that previous communications remain confidential even if static credentials are exposed in the future.

\subsubsection{Insider Attack Resistance:} The proposed protocol withstands insider attacks because a valid session key requires contributions from both parties. Every handshake message is protected by an Ed25519 signature, so even if an attacker steals one long‑term private key ($k_{pr,c}, k_{pr,s}$) they can forge only messages of the participant. In the key agreement, the initial key $k_{init}$ protects the exchange of ephemeral public keys but does not reveal the final secret. The ephemeral key $k_{eph}$ depends on both ephemeral private values, which are deleted immediately after use. Furthermore, the final session key $k_{sym}$is explicitly bound to the nonce $n_r$ through a hash, ensuring freshness and mutual confirmation. As a result, possession of a single long-term key or even knowledge of $k_{init}$ is insufficient to derive $k_{sym}$ preventing an insider from impersonating the other peer or reconstructing session traffic.

\subsubsection{Impersonation Attack Resistance:} The proposed protocol resists impersonation attacks by certificate-based authentication. In the mutual authentication phase, each side proves its identity by presenting a valid certificate, timestamp, and Ed25519 signature. Only the legitimate holders of the private signing keys can generate these values, so an adversary cannot pose as either party. The use of fresh timestamps further ensures that replayed authentication messages are rejected. Since subsequent key exchange depends on ephemeral secrets protected under $k_{init}$, even a party with partial information cannot impersonate the other peer. Finally, explicit key confirmation through a hash that binds the session key to the nonce $n_r$ guarantees that only legitimate peers can establish a valid session key, preventing an adversary from completing the protocol on behalf of another entity.

\subsubsection{Denial of Service Attack Resistance (DoS):} The proposed protocol reduces the impact of DoS attacks by imposing low-cost validation steps before more expensive cryptographic operations are carried out. The validity of cetificate and freshness of timestamps are verified first, ensuring that stale or malformed packets are discarded without further processing. Only after these preliminary checks do the parties derive $k_{init}$ and engage in the ephemeral key exchange. Forged or replayed ciphertexts inevitably fail decryption, tag verification, or final hash comparison, causing the session to terminate without significant computational overhead. In addition, Phase 1 authentication is executed only once to establish long-term trust, while subsequent sessions require only Phase 2 to derive fresh keys. This design minimizes repeated reliance on costly certificate checks and further reduces the susceptibility of constrained devices to denial of service attempts.

\subsection{Comparative Analysis}
In this section, we provide a comparative evaluation between our proposed protocol and related schemes reported previously in the literature. The primary objective is to clearly demonstrate the novelty, strengths, and security improvements achieved by our approach. To facilitate this analysis, we summarize key security characteristics and resistance to common cryptographic attacks in Table ~\ref{tab:comparative}. This table highlights several shortcomings prevalent in existing authentication schemes. For example, the protocol proposed by Garg et al. \cite{garg2019secure} offers limited mutual authentication, relying mainly on simple one-way hash functions, which can be easily compromised. Furthermore, this protocol remains vulnerable to insider attacks due to insufficient internal security measures. Similarly, the scheme proposed by Chaudhry et al. lacks robust mutual authentication and session-key agreement capabilities, making it susceptible to significant threats such as MiTM , insider, and DoS attacks \cite{chaudhry2020securing}. Likewise, the approach by Tanveer et al. does not adequately address mutual authentication and the establishment of session keys and is also vulnerable to DoS attacks \cite{tanveer2023lacp}. The authentication method proposed by Chen et al. \cite{chen2021lightweight} exhibits comparable weaknesses, failing to provide adequate mutual authentication and resistance against session key-related and DoS attacks. Although the protocols presented by Aziz et al. \cite{aziz2018lightweight} and Aghapour et al. \cite{aghapour2020ultra}, ensure basic mutual authentication and replay attack protection, the protocols neglect comprehensive defense against other critical attacks such as insider and DoS attacks and many others.
In contrast to existing schemes, LSEG comprehensively addresses these security gaps. It provides rigorous mutual authentication, secure session-key agreement, and robust protection against insider, impersonation, MiTM, replay, and DoS attacks. Additionally, its lightweight and efficient design significantly reduces computational overhead, making it practical and particularly suitable for resource-constrained smart-grid and IoT deployments.

\begin{table*}[t]
\caption{Comparative Analysis}
\label{tab:comparative}
\centering
\begin{tabular}{p{0.32\linewidth} c c c c c c c}  
 & LSEG & \cite{garg2019secure} & \cite{chaudhry2020securing} & \cite{tanveer2023lacp} & \cite{chen2021lightweight} & \cite{aziz2018lightweight} & \cite{aghapour2020ultra} \\
\toprule

Mutual Authentication & $\checkmark$ & \checkmark & $\times$ & $\times$ & $\times$ & \checkmark & \checkmark  \\ 

Session Key Agreement & \checkmark & \checkmark & \checkmark & $\times$ & $\times$ & $\times$ & $\times$ \\ 

Forward Secrecy & \checkmark & \checkmark & $\times$ & \checkmark & \checkmark & $\times$ & $\times$ \\ 

Resistant Against Man-in-the-Middle Attack & \checkmark & \checkmark & \checkmark & $\times$ & \checkmark & $\times$ & $\times$ \\ 

Resistant Against Replay Attack & \checkmark & \checkmark & \checkmark & \checkmark & \checkmark & \checkmark & \checkmark  \\ 

Resistant Against Insider Attack & \checkmark & $\times$ & $\times$ & \checkmark & \checkmark & $\times$ & $\times$ \\ 

Resistant Against DoS Attack & \checkmark & \checkmark & $\times$ & $\times$ & $\times$ & $\times$ & $\times$ \\ 

Resistant Against Impersonation Attack & \checkmark & \checkmark & \checkmark & \checkmark & \checkmark & $\times$ & \checkmark \\ 

\bottomrule
\end{tabular}
\end{table*}
\section{Results}\label{sec:results}

This section presents the experimental results of the proposed protocol, which were evaluated on both a high-performance desktop system and an embedded device. The desktop system is equipped with an Intel Core i9-14900 CPU, featuring 24 cores (32 threads) and capable of clock speeds reaching up to 5.8 GHz. It operates on Ubuntu $24.04$ LTS and has a 64-bit x86\_64 architecture. In contrast, the embedded setup employs a Raspberry Pi 500, which includes a $2.4GHz$ quad-core 64-bit ARM Cortex-A76 CPU (incorporating cryptography extensions), $8GB$ of LPDDR4X RAM and $32GB$ of Class A2 microSD storage, along with hardware support for H.265 decoding. These platforms facilitate a comprehensive performance evaluation across both general-purpose and resource-constrained environments.

\subsection{Performance Analysis}

\subsubsection{Execution time of each cryptographic primitive:}

Table~\ref{tab:Time in each} shows the execution time for each cryptographic primitive used in the proposed protocol, measured on both the embedded device and the desktop system. 
The generation and verification of certificates constitute the primary sources of cryptographic overhead on both platforms. The generation of certificates takes approximately 2263.27 $\mu$s, while verification requires 2059.41 $\mu$s on the Raspberry Pi. These values are considerably higher compared to the Core i9 system, which completes the same operations in 1,834.85 $\mu$s and 1,431.24 $\mu$s, respectively. The disparity underscores the computational intensity of public-key operations on resource-limited hardware.

Key generation using ECDH exhibits improved efficiency in both systems, with execution times of 197.08 $\mu s$ on the Raspberry Pi and 112.48 $\mu s$ on Core i9. Furthermore, the birational mapping step, which serves to transform or map identities or keys, takes 49.77 $\mu s$ on the Raspberry Pi and 25.93 $\mu$s on the desktop.
The hashing in step 4 takes very little time, approximately 11.05 $\mu s$ and 32.5 $\mu s$ on Core i9 and Raspberry Pi, respectively. The ASCON encryption and decryption primitives are the most efficient, which makes them especially suited to embedded environments. The encryption on the Raspberry Pi is completed in 1.79 $\mu s$ and the decryption in 0.91 $\mu s$, while the core i9 system performs these operations even faster, at 0.94 $\mu s$ and 0.75 $\mu s$, respectively. These results confirm the effectiveness of ASCON as a lightweight cipher, which makes it highly suitable for constrained devices where performance and energy efficiency are critical.

\begin{table}[t]
\centering
\caption{Time utilized by each cryptographic primitive in LSEG ($\mu s$)}
\label{tab:Time in each}
\begin{tabular}{p{0.45\linewidth} c c}
\textbf{Cryptographic Primitive} & \textbf{Raspberry Pi} & \textbf{Core i9} \\
\toprule

Certificate Generation    & 2263.27 & 1834.85 \\
Certificate Verification  & 2059.41 & 1431.24 \\
Key Generation (ECDH)     & 197.08  & 112.48  \\
Birational Mapping        & 49.77   & 25.93   \\
Hashing                   &  32.5   & 11.05   \\
ASCON Encryption          & 1.79    & 0.94    \\
ASCON Decryption          & 0.91    & 0.75    \\ 

\bottomrule
\end{tabular}
\end{table}

\subsubsection{Execution time of each protocol step:}

Table~\ref{tab:scales-ablation} provides a comprehensive breakdown of the execution times for each phase of the protocol, measured on both the Raspberry Pi and the desktop systems. This evaluation categorizes the protocol into two distinct phases: authentication and trust establishment phase and secure session key derivation phase. Each phase is further subdivided into two sub phases; step 1 mutual authentication phases, step 2 key derivation phase; step 3 Ephemeral key derivation phase and step 4 key exchange phase, as shown in Figure \ref{fig_1}.

Step 1 includes the generation and verification of certificates. This phase is identified as the most resource-intensive component of the protocol. Specifically, on the Raspberry Pi, this phase requires 4656.12 $\mu$s, while execution times are notably lower on the desktop client at 3905.53 $\mu$s and the server at 3492.39 $\mu$s. These outcomes are consistent with previous analyses of cryptographic primitives, underscoring the substantial overhead associated with asymmetric operations on embedded devices.

Step 2 involves shared secret derivation using ECDH, reprojection of the derived key from ECDH on the y-axis, concatenation, and finally key derivation using HKDF. This phase significantly reduces the execution time to 474.39 $\mu s$ on the Raspberry Pi and 116.71 $\mu s$ on the desktop for the client, respectively. The server side also takes approximately 121.79 $\mu s$ on a desktop. The first two steps will be used only once to establish the trust between the users, after that, only steps 3 and 4 will be utilized in the communication.

Step 3 includes ephemeral key generation using ECDHE. In this step, the generation of the ephemeral key takes the execution times of 335.10 $\mu s$ on Raspberry Pi, 203.34 $\mu s$ on the desktop client side and 172.34 $\mu s$ on the server. Step 4 includes nonce challenge encryption and decryption, symmetric key generation, and XOR operations, with execution times of 57.83 $\mu s$ on the Raspberry Pi, 66.50 $\mu s$ on the desktop client, and 41.50 $\mu s$ on the server. The minimal latency in this phase highlights the lightweight nature of post-handshake communication and the efficiency of the protocol in secure exchanges. These results confirm the suitability of ASCON as a symmetric cipher, particularly in bandwidth or power-constrained environments.

Overall, the protocol is well suited for both embedded and general-purpose settings. Although certificate verification is computationally expensive , but Phase 1 is executed only once, making the overall protocol lightweight for subsequent communications. The resulting fast communication makes it practical for real-time applications, such as secure smart grid communication.

\begin{table}[t]
\centering
\caption{Time utilized in each step of LSEG ($\mu s$)}
\label{tab:scales-ablation}
\fontsize{8}{11}\selectfont
\renewcommand{\arraystretch}{1.2}
\setlength{\tabcolsep}{3pt} 

\begin{tabular}{p{0.15\linewidth} c c c}
\textbf{Protocol Steps} & \textbf{Raspberry Pi (Client)} & \textbf{Core i9 (Client)} & \textbf{Core i9 (Server)} \\
\toprule

Step 1 & 4656.12 & 3905.53 & 3492.39 \\
Step 2 & 474.39  & 116.71  & 121.79  \\
Step 3 &   335.10 & 203.34  &  172.34 \\
Step 4 &   57.83   & 66.50   & 41.50   \\
\bottomrule
\end{tabular}
\end{table}

\subsection{Communication Cost}

Table \ref{tab:4} presents a comparative analysis of communication overhead, detailing both the total number of messages exchanged and the total number of bits transmitted. In the proposed protocol, two phases are performed. Phase 1 is performed only once, or when authentication and trust need to be re-established. The proposed protocol involves a total of three messages being exchanged, occurring during Step 1, Step 3, and Step 4. Step 2 does not involve data exchange, as it is dedicated solely to internal computation and state updates. The above two steps will be performed once; the subsequent steps are executed repeatedly for each session. In Step 3, the client and the server generated the ephemeral key pairs and exchanged their encrypted public keys $E_1, E_2$ under ASCON128a using the initial key $k_{init}$ generated in Step 2. Each public key measures 256 bits, totaling 512 bits transmitted in this step.

During Step 4, the client transmits an encrypted nonce (denoted as $C_1$) of 128 bits for this phase. The server subsequently provides a second ciphertext ($C_2$), which includes a 128-bit symmetric key that is encrypted under $k_{eph}$ using ASCON128a and authenticated with a 256-bit hash, totaling 384 bits (comprising a 128-bit encrypted key and a 256-bit hash). Consequently, the overall communication cost for the proposed protocol is 1024 bits across the two message exchanges.
Since Phase 1 is executed only once, recurring Phase 2 will cost only 1024 bits (512 + 512 bits) across two message exchanges, as summarized in Table \ref{tab:4}.This demonstrates that the proposed protocol is lightweight and communication efficient.

The proposed protocol LSEG exhibits an efficient communication profile compared to existing schemes. Although it transmits more bits than \cite{aghapour2020ultra} (944 bits in three messages), which performs mutual authentication and is resistant to replay and impersonation attacks, but does not provide forward secrecy or session key agreement, LSEG also shows greater efficiency than \cite{tanveer2023lacp} (1206 bits in two messages), which lacks mutual authentication and session key agreement, and is also vulnerable to MiTM and DoS attacks. Additionally, it remains more efficient than \cite{mahmood2016lightweight} and \cite{wu2011fault}, which incur 4352 and 3648 bits, respectively. Moreover, it offers stronger security guarantees than \cite{chen2021lightweight} (1344 bits), which lacks mutual authentication and session key agreement, while maintaining comparable communication complexity.

LSEG strikes a reasonable balance between security guarantees and communication efficiency, positioning it as an optimal solution for constrained environments, such as smart grid edge nodes.

\begin{table}[t]
\centering
\caption{Comparison in communication cost (bits)}
\label{tab:4}
\begin{tabular}{p{0.25\linewidth} c c}
\textbf{Schemes} & \textbf{Messages Exchanged} & \textbf{Bits Exchanged} \\
\toprule

LSEG & 2 & 1024 \\
\cite{mahmood2016lightweight} & 2 & 4352 \\
\cite{tanveer2023lacp} & 2 & 1206 \\
\cite{chen2021lightweight} & 2 & 1344 \\
\cite{aghapour2020ultra} & 3 & 944 \\
\cite{wu2011fault} & 4 & 3648 \\

\bottomrule
\end{tabular}
\end{table}
\section{Conclusion}\label{sec:conclusion}

This paper has presented a lightweight, certificate-based authentication, and secure communication protocol designed for smart grid environments. The protocol employs ASCON128a for authenticated encryption, ensuring data confidentiality and integrity with minimal computational overhead. Formal and informal analyses confirm the resilience of the protocol against various cryptographic threats, including man-in-the-middle, replay, impersonation, and insider attacks, while providing forward secrecy. The experimental evaluation shows that the protocol performs efficiently in both resource-constrained and high-performance environments. The total execution time of the protocol on the client side is approximately 5.5 ms on a Raspberry Pi and 4.3 ms on a desktop, respectively. The communication overhead is limited to 1024 bits across two message exchanges, outperforming several existing protocols in terms of both efficiency and security guarantees. These findings demonstrate that the proposed protocol provides a scalable and practical solution to secure communication in smart grid infrastructures. 

Future work may include integrating post-quantum cryptographic primitives and hardware-based trust anchors, such as a Trusted Platform Module (TPM), to further enhance resilience and long-term security.

%
%
\bibliographystyle{splncs04}
\bibliography{references}

@inproceedings{zyskind2015decentralizing,
  title={Decentralizing privacy: Using blockchain to protect personal data},
  author={Zyskind, Guy and Nathan, Oz and others},
  booktitle={2015 IEEE security and privacy workshops},
  pages={180--184},
  year={2015},
  organization={IEEE}
}

@article{mcdaniel2009security,
  title={Security and privacy challenges in the smart grid},
  author={McDaniel, Patrick and McLaughlin, Stephen},
  journal={IEEE security \& privacy},
  volume={7},
  number={3},
  pages={75--77},
  year={2009},
  publisher={IEEE}
}

@article{karale2021challenges,
  title={The challenges of IoT addressing security, ethics, privacy, and laws},
  author={Karale, Ashwin},
  journal={Internet of Things},
  volume={15},
  pages={100420},
  year={2021},
  publisher={Elsevier}
}

@phdthesis{sandeep2006elliptic,
  title={Elliptic curve cryptography for constrained devices},
  author={Sandeep, S},
  year={2006},
  school={PhD Dissertation}
}

@inproceedings{won2018decentralized,
  title={Decentralized public key infrastructure for internet-of-things},
  author={Won, Jongho and Singla, Ankush and Bertino, Elisa and Bollella, Greg},
  booktitle={MILCOM 2018-2018 IEEE Military Communications Conference (MILCOM)},
  pages={907--913},
  year={2018},
  organization={IEEE}
}

@article{garg2019secure,
  title={Secure and lightweight authentication scheme for smart metering infrastructure in smart grid},
  author={Garg, Sahil and Kaur, Kuljeet and Kaddoum, Georges and Rodrigues, Joel JPC and Guizani, Mohsen},
  journal={IEEE Transactions on Industrial Informatics},
  volume={16},
  number={5},
  pages={3548--3557},
  year={2019},
  publisher={IEEE}
}

@book{hartshorne2013algebraic,
  title={Algebraic geometry},
  author={Hartshorne, Robin},
  volume={52},
  year={2013},
  publisher={Springer Science \& Business Media}
}

@article{bernstein2012high,
  title={High-speed high-security signatures},
  author={Bernstein, Daniel J and Duif, Niels and Lange, Tanja and Schwabe, Peter and Yang, Bo-Yin},
  journal={Journal of cryptographic engineering},
  volume={2},
  number={2},
  pages={77--89},
  year={2012},
  publisher={Springer}
}

@inproceedings{krawczyk2010cryptographic,
  title={Cryptographic extraction and key derivation: The HKDF scheme},
  author={Krawczyk, Hugo},
  booktitle={Annual Cryptology Conference},
  pages={631--648},
  year={2010},
  organization={Springer}
}

@inproceedings{gilchrist2008secure,
  title={Secure authentication for DNP3},
  author={Gilchrist, Grant},
  booktitle={2008 IEEE Power and Energy Society General Meeting-Conversion and Delivery of Electrical Energy in the 21st Century},
  pages={1--3},
  year={2008},
  organization={IEEE}
}

@article{iec200762351,
  title={62351-3:‘Power systems management and associated information exchange--data and communications security--part 3: communication network and system security--profiles including TCP},
  author={IEC, TS},
  journal={IP’(International Electrotechnical Commission, Geneva, Switzerland, 2007)},
  year={2007}
}

@misc{dlms2019green,
  title={Green book, DLMS/COSEM Architecture and Protocols},
  author={DLMS User Association and others},
  year={2019},
  publisher={Ed}
}

@article{ieee2018ieee,
  title={Ieee standard for smart energy profile application protocol},
  author={IEEE Standards Association and others},
  journal={Accessed: Nov},
  volume={20},
  number={2023},
  pages={20305--2018},
  year={2018}
}

@article{case2016analysis,
  title={Analysis of the cyber attack on the Ukrainian power grid},
  author={Case, Defense Use},
  journal={Electricity information sharing and analysis center (E-ISAC)},
  volume={388},
  number={1-29},
  pages={3},
  year={2016},
  publisher={Washington, DC}
}

@article{ten2010cybersecurity,
  title={Cybersecurity for critical infrastructures: Attack and defense modeling},
  author={Ten, Chee-Wooi and Manimaran, Govindarasu and Liu, Chen-Ching},
  journal={IEEE Transactions on Systems, Man, and Cybernetics-Part A: Systems and Humans},
  volume={40},
  number={4},
  pages={853--865},
  year={2010},
  publisher={IEEE}
}

@article{sridhar2011cyber,
  title={Cyber--physical system security for the electric power grid},
  author={Sridhar, Siddharth and Hahn, Adam and Govindarasu, Manimaran},
  journal={Proceedings of the IEEE},
  volume={100},
  number={1},
  pages={210--224},
  year={2011},
  publisher={IEEE}
}

@article{rivest1978method,
  title={A method for obtaining digital signatures and public-key cryptosystems},
  author={Rivest, Ronald L and Shamir, Adi and Adleman, Leonard},
  journal={Communications of the ACM},
  volume={21},
  number={2},
  pages={120--126},
  year={1978},
  publisher={ACM New York, NY, USA}
}

@article{koblitz1987elliptic,
  title={Elliptic curve cryptosystems},
  author={Koblitz, Neal},
  journal={Mathematics of computation},
  volume={48},
  number={177},
  pages={203--209},
  year={1987}
}

@techreport{cooper2008internet,
  title={Internet X. 509 public key infrastructure certificate and certificate revocation list (CRL) profile},
  author={Cooper, David and Santesson, Stefan and Farrell, Stephen and Boeyen, Sharon and Housley, Russell and Polk, William},
  year={2008}
}

@article{aziz2018lightweight,
  title={A lightweight scheme to authenticate and secure the communication in smart grids},
  author={Aziz, Israa T and Jin, Hai and Abdulqadder, Ihsan H and Hussien, Zaid Alaa and Abduljabbar, Zaid Ameen and Flaih, Firas MF},
  journal={Applied Sciences},
  volume={8},
  number={9},
  pages={1508},
  year={2018},
  publisher={MDPI}

}

@article{mahmood2016lightweight,
  title={A lightweight message authentication scheme for smart grid communications in power sector},
  author={Mahmood, Khalid and Chaudhry, Shehzad Ashraf and Naqvi, Husnain and Shon, Taeshik and Ahmad, Hafiz Farooq},
  journal={Computers \& Electrical Engineering},
  volume={52},
  pages={114--124},
  year={2016},
  publisher={Elsevier}
}

@article{li2013enhanced,
  title={An enhanced smart card based remote user password authentication scheme},
  author={Li, Xiong and Niu, Jianwei and Khan, Muhammad Khurram and Liao, Junguo},
  journal={Journal of Network and Computer Applications},
  volume={36},
  number={5},
  pages={1365--1371},
  year={2013},
  publisher={Elsevier}
}

@article{gope2018lightweight,
  title={Lightweight and privacy-preserving two-factor authentication scheme for IoT devices},
  author={Gope, Prosanta and Sikdar, Biplab},
  journal={IEEE Internet of Things Journal},
  volume={6},
  number={1},
  pages={580--589},
  year={2018},
  publisher={IEEE}
}

@article{challa2018efficient,
  title={An efficient ECC-based provably secure three-factor user authentication and key agreement protocol for wireless healthcare sensor networks},
  author={Challa, Sravani and Das, Ashok Kumar and Odelu, Vanga and Kumar, Neeraj and Kumari, Saru and Khan, Muhammad Khurram and Vasilakos, Athanasios V},
  journal={Computers \& Electrical Engineering},
  volume={69},
  pages={534--554},
  year={2018},
  publisher={Elsevier}
}

@article{abbasinezhad2017ultra,
  title={An ultra-lightweight and secure scheme for communications of smart meters and neighborhood gateways by utilization of an ARM Cortex-M microcontroller},
  author={Abbasinezhad-Mood, Dariush and Nikooghadam, Morteza},
  journal={IEEE Transactions on Smart Grid},
  volume={9},
  number={6},
  pages={6194--6205},
  year={2017},
  publisher={IEEE}
}

@article{baghestani2022lightweight,
  title={Lightweight authenticated key agreement for smart metering in smart grid},
  author={Baghestani, Seyed Hamid and Moazami, Farokhlagha and Tahavori, Mahdi},
  journal={IEEE Systems Journal},
  volume={16},
  number={3},
  pages={4983--4991},
  year={2022},
  publisher={IEEE}
}

@article{chen2021lightweight,
  title={Lightweight authentication protocol in edge-based smart grid environment},
  author={Chen, Chien-Ming and Chen, Lili and Huang, Yanyu and Kumar, Sachin and Wu, Jimmy Ming-Tai},
  journal={EURASIP Journal on Wireless Communications and Networking},
  volume={2021},
  number={1},
  pages={68},
  year={2021},
  publisher={Springer}
}

@article{tanveer2023lacp,
  title={LACP-SG: Lightweight authentication protocol for smart grids},
  author={Tanveer, Muhammad and Alasmary, Hisham},
  journal={Sensors},
  volume={23},
  number={4},
  pages={2309},
  year={2023},
  publisher={MDPI}
}

@article{chaudhry2020securing,
  title={Securing demand response management: A certificate-based access control in smart grid edge computing infrastructure},
  author={Chaudhry, Shehzad Ashraf and Alhakami, Hosam and Baz, Abdullah and Al-Turjman, Fadi},
  journal={IEEE Access},
  volume={8},
  pages={101235--101243},
  year={2020},
  publisher={IEEE}
}

@article{aghapour2020ultra,
  title={An ultra-lightweight and provably secure broadcast authentication protocol for smart grid communications},
  author={Aghapour, Saeed and Kaveh, Masoud and Mart{\'\i}n, Diego and Mosavi, Mohammad Reza},
  journal={IEEE Access},
  volume={8},
  pages={125477--125487},
  year={2020},
  publisher={IEEE}
}

@article{wu2011fault,
  title={Fault-tolerant and scalable key management for smart grid},
  author={Wu, Dapeng and Zhou, Chi},
  journal={IEEE Transactions on Smart Grid},
  volume={2},
  number={2},
  pages={375--381},
  year={2011},
  publisher={IEEE}
}

@ARTICLE{MunirEdgeAIAESM2021,
  author={Munir, Arslan and Blasch, Erik and Kwon, Jisu and Kong, Joonho and Aved, Alexander},
  journal="{IEEE Aerospace and Electronic Systems Magazine}", 
  title="{Artificial Intelligence and Data Fusion at the Edge}", 
  year={2021},
  volume={36},
  number={7},
  pages={62--78},
  doi={10.1109/MAES.2020.3043072}
}

\end{document}